\documentclass[11pt, a4paper]{amsart}

\usepackage{color}

\usepackage{latexsym}

\usepackage{amssymb}
\usepackage[centertags]{amsmath}
\usepackage{verbatim}
\usepackage{epic, eepic}

\newtheorem{theorem}{Theorem}
\newtheorem{lemma}[theorem]{Lemma}
\newtheorem{claim}[theorem]{Claim}
\newtheorem{remark}[theorem]{Remark}

\numberwithin{theorem}{section}

\newcommand{\ci}[1]{_{{}_{\scriptstyle #1}}}

\renewcommand{\phi}{\varphi}

\newcommand{\N}{\mathbb N}

\newcommand{\la}{\lambda}

\newcommand{\E}{\mathcal E}

\renewcommand{\epsilon}{\varepsilon}

\def\done{{1\hskip-2.5pt{\rm l}}}

\renewcommand{\le}{\leqslant}
\renewcommand{\ge}{\geqslant}

\newcommand{\R}{\mathbb R}
\newcommand{\C}{\mathbb C}

\newcommand{\cZ}{\mathcal Z}

\newcommand{\T}{\mathbb T}

\title
[Correlation functions for random complex zeroes]
{Correlation functions for random complex zeroes:
strong clustering and local universality}

\author{F. Nazarov}
\address{F.N.: Mathematics Department \\
University of Wisconsin-Madison \\
480 Lincoln Dr., Madison WI 53706 \\
USA}

\email{nazarov@math.wisc.edu}

\author{M. Sodin}
\address{M.S.: School of Mathematical Sciences\\
Tel Aviv University\\
Tel Aviv 69978\\
Israel}

\email{sodin@post.tau.ac.il}

\thanks{F.N. is partially supported by
the National Science Foundation, DMS grant 0501067. M.S. is
partially supported by the Israel Science Foundation of the Israel
Academy of Sciences and Humanities, grant 171/07}

\begin{document}

\date{May 22, 2010}

\begin{abstract}

We prove strong clustering of $k$-point correlation functions of zeroes
of Gaussian Entire Functions. In the course of the proof, we also obtain
universal local bounds for $k$-point functions of zeroes of arbitrary
nondegenerate Gaussian analytic functions.

In the second part of the paper,
we show that strong clustering yields the asymptotic normality of fluctuations
of some linear statistics of zeroes of Gaussian Entire Functions, in
particular, of the number of zeroes in measurable domains of large
area. This complements our recent results
from the paper ``Fluctuations in random complex zeroes''.

\end{abstract}

\maketitle

\section{Introduction}\label{sect_Intro}

Consider the Gaussian entire function (G.E.F., for short)
$\displaystyle F(z) = \sum_{j\ge 0} \zeta_j \frac{z^j}{\sqrt{j!}} $,
where $\zeta_j$ are standard independent Gaussian complex coefficients;
that is, the density of the probability distribution of
$\zeta_j$ with respect to the Lebesgue measure on the complex plane is
$\tfrac1{\pi} e^{-|\zeta|^2}$.
A remarkable feature of the random zero set $\cZ_F=F^{-1}\{0\}$ is its distribution
invariance with respect to the isometries of the complex plane \cite{HKPV, NS1, ST}.
It's easy to compute the covariance function of $F$:
\[
\E \bigl\{ F(z) \overline{F(w)} \bigr\} = \sum_{j\ge 0} \frac{z^j\, \overline{w}\,^j}{j!}
= e^{z\overline{w}}\,.
\]
Hence, after normalization, the covariance equals
$ e^{z\overline{w}-\frac12|z|^2-\frac12 |w|^2} = e^{{\rm i}{\rm Im}(z\overline{w}) - \frac12 |z-w|^2}$,
which decays very fast when $|z-w|$ grows. This hints at the almost independence
of random zeroes at large distances. One of convenient ways to formalize
the almost independence of a zero process at large distances
is based on {\em clustering} of its $k$-point functions, cf.~\cite[\S~4.4]{Ruelle}.
In this paper, we will develop this idea for zeroes of G.E.F.'s, and then
will apply it to prove the asymptotic normality of fluctuations
of some linear statistics of zeroes of Gaussian Entire Functions.
We note that there is a very different approach to almost independence of random
complex zeroes on large distances that proved to be useful in~\cite{NS, NSV1, NSV2}.

\subsection{$k$-point functions}
The $k$-point functions express correlations within $k$-point
subsets of the point process. It is customary in statistical mechanics
to describe point processes by the properties of their $k$-point correlation functions.

The $k$-point function $\rho = \rho_k$ of a random zero process $\mathcal Z$ on $\C$
is a symmetric function
\[
\rho\colon \bigl\{Z\!=\!(z_1, ..., z_k)\!\in\!\C^k\colon z_i\ne z_j, {\ \rm for\ } i\ne j  \bigr\} \to \R_+
\]
defined by the formula
\[
\E \bigl\{ \prod_{1\le i \le k} \# \left( \mathcal Z \cap B_i \right)
\bigr\}
= \int_{B_1\times ... \times B_k} \rho (z_1, ..., z_k)\, {\rm d}A(z_1) ... {\rm d} A(z_k)\,,
\]
for any family of mutually disjoint bounded Borel sets $B_1$, ..., $B_k$ in $\C$.
Here $A$ is the Lebesgue measure on the complex plane.
The $k$-point function of zeroes of a Gaussian analytic function $f$ exists,
provided that for any $z_1, ..., z_k\in\C$ with $z_i\ne z_j$ for $i\ne j$,
the random variables $f(z_1)$, ..., $f(z_k)$ are linearly
independent~\cite[Corollary~3.4.2]{HKPV}.
It is not difficult to show (see the beginning of Section~\ref{sect-weak_univ})
that for G.E.F.'s, this condition holds for each $k\in \N$.

\subsection{Main results}
The first result treats the local behaviour of $k$-point functions.
It appears that for a wide class of non-degenerate Gaussian
analytic functions, the $k$-point functions of their zeroes
exhibit universal local repulsion when some of the variables
$z_1, ..., z_k$ approach each other.

Recall that a Gaussian analytic function (G.a.f., for short) $ f(z) $ in a plane domain
$ G\subseteq \C $ is the sum
\begin{equation}\label{eq-10}
f(z) = \sum_n \zeta_n f_n (z)
\end{equation}
of analytic functions $ f_n(z) $ such that
\[
\sum_n |f_n(z)|^2 < \infty \quad \text{locally  uniformly  on } G,
\]
with independent standard complex Gaussian coefficients
$\zeta_n$.

We postpone until the beginning of Section~\ref{sect-weak_univ}
the technical definition of $d$-degeneracy, which we use
in the assumptions of the next theorem. Here, we only mention that
G.a.f.'s with ``deterministic zeroes'' (that is, $f(z_0)=0$ a.s.,
for some $z_0\in G$) are $1$-degenerate. G.a.f.'s  such that the random variables
$ f(z_1), ..., f(z_k) $ are linearly dependent for some $z_1, ..., z_k \in G$,
are $k$-degenerate, and G.a.f.'s for which the random variables
$ f(z_1), f'(z_1), ..., f(z_k), f'(z_k) $ are linearly dependent are
$2k$-degenerate. We also mention that Gaussian Taylor series (either infinite, or finite)
\[
f(z) = \sum_{n\ge 0} \zeta_n c_n z^n
\]
are $d$-nondegenerate,  provided that $c_0, c_1, ..., c_{d-1} \ne 0$.
In particular, the G.E.F. is $d$-nondegenerate for every positive integer $d$.

\begin{theorem}\label{thm-weak-univ}
Let $f$ be a $2k$-nondegenerate G.a.f. in a domain $G$, let $\rho_f $ be a $k$-point
function of zeroes of $f$, and let $K\subset G$ be a compact set.
Then there exists a positive constant $C=C(k, f, K)$ such that,
for any configuration of pairwise distinct points $z_1, ..., z_k\in K$,
\[
C^{-1} \prod_{i<j} |z_i-z_j|^2 \le \rho_f (z_1, ..., z_k)
\le C \prod_{i<j} |z_i-z_j|^2\,.
\]
\end{theorem}

The next result is a clustering property of zeroes of G.E.F.'s.
It says that if the
variables in $\C^k$ can be split into two groups located far from each other,
then the function $\rho_k$ almost equals the product of the corresponding
factors. This property is another manifestation of almost independence of
points of the process at large distances.

For a non-empty subset
$I = \left\{ i_1, ..., i_\ell \right\} \subset \left\{ 1, 2, ..., k \right\}$,
we set $Z_I = \left\{ z_{i_1}, ..., z_{i_\ell} \right\}$. We denote by
\[ d(Z_I, Z_J) = \inf_{i\in I, j\in J} |z_i-z_j| \] the distance
between the configurations $ Z_I $ and $ Z_J $.
\begin{theorem}\label{thm_main*}
For each $k\ge 2$, there exist positive constants $C_k$ and $\Delta_k$ such that for
each configuration $Z$ of size $k$ and each partition of the set of indices
$\left\{ 1, 2, ..., k \right\}$ into two non-empty
subsets $ I $ and $ J $ with $ d(Z_I, Z_J) \ge 2\Delta_k $, one has
\begin{equation}\label{eq:cluster}
1 - \delta
\le \frac{ \rho (Z) }{ \rho (Z_I) \rho (Z_J)}
\le 1 + \delta
\quad{\rm with} \quad
\delta =  C_k e^{-\frac12 (d(Z_I, Z_J)-\Delta_k)^2}\,.
\end{equation}
\end{theorem}

Combining Theorems~\ref{thm-weak-univ} and~\ref{thm_main*}, and taking into
account the translation invariance of the random zero process $\mathcal Z_F$, we obtain
a uniform estimate for $\rho_k$ valid in the whole $\C^k$:
\begin{theorem}\label{thm-multipl} For each $k\ge 1$,
there exists a positive constant $C_k$ such that for each configuration
$(z_1, ..., z_k)$,
\[
C_k^{-1} \prod_{i<j} \ell (|z_i-z_j|)
\le \rho (z_1, ..., z_k) \le C_k \prod_{i<j} \ell(|z_i-z_j|)\,,
\]
where $\ell (t) = \min(t^2, 1)$.
\end{theorem}
To get Theorem~\ref{thm-multipl}, we use induction on $k$.
For $k=1$ the result is obvious. Now, given an integer $m\ge 2$, suppose that Theorem~\ref{thm-multipl}
has been proven for $k$-point functions with $k\le m$. To prove it for $m+1$-point functions,
we consider two cases. First, suppose that the configuration  of points
$\{z_1, ..., z_{m+1}\}$ cannot be split into two groups lying at distance $2\Delta_{m+1}$
or more from each other. Then the result follows from the local bounds in Theorem~\ref{thm-weak-univ}
and translation invariance.
In the other case, the result follows from Theorem~\ref{thm_main*} and
the inductive assumption. \hfill $\Box$

\medskip
Note that Theorem~\ref{thm-multipl} yields that the $k$-point functions $\rho_k$
are uniformly bounded on $\C^k$ by constants depending on $k$. This observation
immediately yields
the additive version of the clustering property:
\begin{theorem}\label{thm_main}
For each $k\ge 2$, there exist positive constants $C_k$ and $\Delta_k$ such that for
each configuration $Z$ of size $k$ and each partition of the set of indices
$\left\{ 1, 2, ..., k \right\}$ into two non-empty
subsets $ I $ and $ J $ with $ d(Z_I, Z_J) \ge 2\Delta_k $,  one has
\begin{equation}\label{eq_clustering}
\left| \rho (Z) - \rho (Z_I) \rho (Z_J) \right|
\le C_k e^{- \frac12 \left( d(Z_I, Z_J) - \Delta_k \right)^2}\,.
\end{equation}
\end{theorem}

\medskip
The  proofs of Theorems~\ref{thm-weak-univ} and~\ref{thm_main*}
start with the classical Kac-Rice-Hammersley  formula~\cite[Chapter~3]{HKPV}:
\begin{equation}\label{eq:KRH*}
\rho_f (z_1, ..., z_k)
= \int_{\C^k} |\eta_1|^2 ... |\eta_k|^2\, \mathcal D_f(\eta'; z_1, ..., z_k)\,
{\rm d} m(\eta_1) ... {\rm d} m(\eta_k),
\end{equation}
where $\mathcal D_f(\, \cdot \,; z_1, ..., z_k)$
is the density of the joint probability distribution of the random variables
\begin{equation}\label{eq-20*}
f(z_1),\, f'(z_1),\, ...\,, f(z_k),\, f'(z_k)\,,
\end{equation}
and $\eta' = \left( 0, \eta_1, ..., 0, \eta_k \right)^{\tt T}$
is a vector in $\C^{2k}$.
Since the random variables~\eqref{eq-20*} are complex Gaussian, one can rewrite
the right-hand side of~\eqref{eq:KRH*} in a more explicit form
\begin{equation}\label{eq-30*}
\rho_f (z_1, ..., z_k) = \frac1{\pi^{2k} \det \Gamma_f }
\int_{\C^k} |\eta_1|^2 ... |\eta_k|^2  e^{-\frac12 \langle \Gamma_f^{-1} \eta', \eta' \rangle}
{\rm d}m(\eta_1) ... {\rm d} m(\eta_k),
\end{equation}
where $\Gamma_f = \Gamma_f (z_1, ..., z_k)$ is the covariance matrix of the random variables~\eqref{eq-20}.
We consider the linear functionals
\[
L f = \sum_{j=1}^k \left[ \alpha_j f(z_j) + \beta_j f'(z_j) \right]
= \frac1{2\pi{\rm i}} \int_{\gamma} f(z) r^L(z)\, {\rm d}z,
\]
where
\[
r^L(z) = \sum_{j=1}^k \left[ \frac{\alpha_j}{z-z_j} + \frac{\beta_j}{(z-z_j)^2} \right],
\]
and $\gamma \subset K$ is a smooth contour that bounds a domain $G'\subset K$ that contains the
points $z_1, ..., z_k$. Then we observe that for every vector
$\delta = \left( \alpha_1, \beta_1, ... , \alpha_k, \beta_k \right)^{\tt T} $
in $\C^{2k}$, we have \[ \langle \Gamma_f \delta, \delta \rangle = \E | Lf |^2\,.\]
This observation allows us to estimate the matrix $\Gamma_f^{-1}$, and hence
the integral on the right-hand side of~\eqref{eq-30*},
using some simple tools from the theory of analytic functions of
one complex variable, and thus avoiding formidable expressions for the Gaussian
integrals on the right-hand side of~\eqref{eq-30*} that involve quotients of
large determinants and permanents.

We note that in \cite{BSZ}, Bleher, Shiffman, and Zelditch estimated these
large determinants and permanents, and proved that if the points $z_i$
are well separated from each other, i.e.,
$\displaystyle \min_{i\ne j}|z_i-z_j|\ge \eta >0$, then some estimate similar
to~\eqref{eq_clustering} holds with a factor $C(k, \eta)$ on the right-hand side.
Unfortunately, in this form their result is difficult to apply. For instance, it does not
yield boundedness of the $k$-point functions on the whole $\C^k$, and we could not
use it for the proof of the asymptotic normality of some linear statistics, see
Theorem~\ref{thm_norm} below.

\subsection{Clustering of $k$-point functions and asymptotic normality of linear
statistics}
Various ways to derive the asymptotic normality of fluctuations of the number
of random points in large volumes from clustering
of all $k$-point functions are known is statistical mechanics.
Our next result is a variation on this theme. It pertains to arbitrary point processes
$\mathcal Z$ on the plane with clustering $k$-point functions.

Let $h$ be a non-zero bounded measurable function with compact support.
For the scaled linear statistics $ n(R; h) = \sum_{a\in\mathcal Z } h\bigl( \frac{a}{R} \bigr) $
of the point process $\mathcal Z$, we set
\begin{multline*}
\overline{n}(R; h) = n(R; h) - \E \{ n(R; h) \}, \quad
\sigma (R; h)^2 = \E \bigl\{ \overline{n}(R; h)^2 \bigr\}, \\
\qquad \text{and} \quad n^*(R; h) = \frac{\overline{n}(R; h)}{\sigma (R; h)}\,. \qquad
\end{multline*}
In what follows, we consider the functions $h$ with $ \sigma (R; h)$
growing as a positive power of $R$, i.e., satisfying
\begin{equation}\label{eq_power_growth}
\sigma (R; h) \ge R^\delta
\end{equation}
for some $\delta>0$ and for all sufficiently big $R$.
Later, we will see that this holds when $h$ is an indicator function of an arbitrary
bounded measurable set of positive area (Lemma~\ref{lemma_var-growth} below).

We call the function $\phi\colon (0, \infty) \to (0, \infty) $
{\em fast decreasing} if it decreases and
for each positive $m$,
$\displaystyle \lim_{x\to \infty} x^m \phi (x) = 0$.
We say that {\em the $k$-point functions
$\rho$ of a point process $\mathcal Z$ are clustering} if
there exits a fast decreasing function $\phi$ such that for each $k\ge 2$
and for each partition of the set of indices $\{1, 2, ..., k\}$ into non-empty disjoint subsets
$I$ and $J$, one has
\begin{equation}\label{eq_def_cluster}
\left| \rho (Z) - \rho(Z_I)\rho(Z_J) \right| \le C_k \phi (c_k d(Z_I, Z_J))\,.
\end{equation}
We say that the $k$-point functions are bounded, if
$ \sup_{\C^k} \rho_k < \infty $.

\begin{theorem}\label{thm_norm}
Suppose that the $k$-point correlation functions $\rho_k$ of a
point process $\mathcal Z$ are
clustering and bounded. Then
for each bounded measurable compactly supported function $h$ with
$ \sigma (R; h) $ growing as a positive power of $R$,
the normalized linear statistics $ n^*(R; h) $
converge in distribution to the standard Gaussian law as $R\to\infty$.
\end{theorem}

The proof of this theorem is based on estimates for cumulants
that follow from clustering. A similar approach was used by Malyshev in \cite{Mal}, and by
Martin and Yal\c{c}in in \cite{MY}.

We note that there is
a counterpart of Theorem~\ref{thm_norm} for arbitrary determinantal point processes, namely,
a theorem of Soshnikov. In~\cite{Soshn}, he proved that for arbitrary determinantal point
processes, the fluctuations of linear statistics associated with compactly supported bounded positive
functions $h$ are asymptotically normal if the variance grows
at least as a positive power of the expectation. His proof is based on peculiar combinatorial
identities for the cumulants of linear statistics that are a special feature of
determinantal point processes, and is quite different from the one of Theorem~\ref{thm_norm}.

\subsection{Lower bounds for the variance of linear statistics}
In order to apply Theorem~\ref{thm_norm}, we need to know that the variance
of the linear statistics $n(R; h)$ grows as a positive power of $R$.
An example of such statistics  is the number of points of the process $\mathcal Z$
in a bounded set $E$ of positive measure dilated $R$ times.

We assume that $ \mathcal Z $
is a {\em translation-invariant} point process on the plane with one- and two-point correlation
functions and
that the mean number of points of the process $ \mathcal Z $ per unit area equals $1$;
that is, $ \rho_1 \equiv 1$ on $\R^2$. We also assume that
the $2$-point function of the point process $\mathcal Z $ satisfies
\begin{equation}\label{eq_weak_cluster}
\rho(z_1, z_2) = r(z_1-z_2) \qquad with \quad r-1\in L^1(\R^2)
\end{equation}
which is a weak form of clustering.
For a set $E$, we denote by $\done_E$ its indicator function.

\begin{lemma}\label{lemma_var-growth} Let $ \mathcal Z $ be a translation-invariant
random point process on $\R^2$ with the $2$-point function satisfying
\eqref{eq_weak_cluster}.
Then there exists a numerical constant $c>0$ such
that for each bounded measurable set $E\subset\R^2$ of positive
area $A(E)$, one has
\[
\sigma (R, \done_E)^2 \ge c \min\left\{ A(E)R^2,  \sqrt{A(E)}\,R \right\}\,,
\qquad 0<R<\infty\,.
\]
\end{lemma}

Note that we do not impose any smoothness assumption on the boundary of
the set $E$. The proof of Lemma~\ref{lemma_var-growth} is
based on a simple observation:

\begin{lemma}\label{lemma-simple}
Suppose $ \mathcal Z $ is a translation-invariant
random point process  on $\R^2$ with the $2$-point function satisfying
\eqref{eq_weak_cluster}. Then there exists a numerical constant $C>0$ such that
for every function $h\in (L^1\cap L^2) (\R^2)$ and every $R>0$,
\begin{equation}\label{eq_variance}
\sigma (R, h)^2 \ge \frac{R^2}2\, \int_{|\xi|\ge CR} |\widehat{h}(\xi)|^2 \, {\rm d} A(\xi) \,,
\end{equation}
where
\[
\widehat{h}(\xi) = \int_{\R^2} h(x)e^{-2\pi {\rm i} x\cdot \xi}\, {\rm d} A(x)
\]
is the Fourier transform of $h$.
\end{lemma}

Estimate~\eqref{eq_variance} reduces lower bounds for $\sigma (R, h)$ to
estimates for the tail of the integral on the right-hand side of~\eqref{eq_variance}.

\subsection{Related results on asymptotic normality of fluctuations of linear statistics
of random complex zeroes} Theorem~\ref{thm_norm} complements
our recent results~\cite{NS} on fluctuations in
the random complex zeroes $\mathcal Z_F$ obtained by a completely different technique.
Therein, we showed that if $ \mathcal Z_F$ is a zero set of a G.E.F. $F(z)$, then the following
hold:

\smallskip\par\noindent (i) If $ h $ is a $C^\alpha_0$-function with $\alpha>1$, then the
fluctuations of $ n(R, h) $ are asymptotically normal.\footnote{This result was preceded by
yet different proofs in~\cite{ST, Tsirel} of the asymptotic normality in the case of $C^2_0$-functions $h$.}

\smallskip\par\noindent (ii) If $ h $ is a $C^\alpha_0$ function with $0<\alpha\le 1$, and if
for some $\epsilon>0$, we have $\sigma (R, h) \ge R^{-\alpha+\epsilon}$, then the fluctuations
of $ n(R, h) $ are asymptotically normal as well.

\smallskip\par\noindent (iii) For each $0<\alpha<1$, there are $C_0^\alpha$-functions $h$
with abnormal fluctuations of linear statistics $ n(R, h) $.

\subsection*{Acknowledgement} We thank Boris Tsirelson for very useful discussions.

\section{Local universality of $k$-point functions. Proof of Theorem~\ref{thm-weak-univ}}
\label{sect-weak_univ}

Let $G\subseteq \C$ be a plane domain. We fix a compact set $K$
and a bounded domain $G_1$ with a smooth boundary $\gamma = \partial G_1$ such that $K\subset G_1$
and $\overline{G_1}\subset G$. We consider linear functionals
\begin{equation}\label{eq-50}
L g = \sum_{j=1}^k a_j g^{(m_j)}(z_j)
\end{equation}
with $z_1, ..., z_k \in K$, acting on functions $g$ analytic in $G$.
Then for every function $g$ analytic in $G$ and for every functional $L$ as above,
we have
\begin{equation}\label{eq-45}
L g = \int_\gamma g(z) r^L (z)\, {\rm d}z \qquad
\text{with}\quad r^L(z) = \frac1{2\pi {\rm i}} \sum_{j=1}^k \frac{a_j m_j!}{(z-z_j)^{m_j+1}}\,.
\end{equation}
Note that $r^L$ is a rational function vanishing at infinity. We put
\[ \text{rank}(L) \stackrel{\text{def}}= \text{deg}(r^L)\,. \]
Then we say that a G.a.f. $f(z)$ in $G$ is $d$-degenerate, if there exists a functional
$L$ of rank at most $d$ such that
\begin{equation}\label{eq-60}
Lf = 0 \qquad \text{almost surely}.
\end{equation}
Otherwise, we say that $f$ is {\em $d$-nondegenerate}. For instance, G.a.f.'s with
``deterministic zeroes'' (that is, $f(z_0)=0$ a.s., for some $z_0\in G$) are $1$-degenerate.
G.a.f.'s  such that the random variables
$ f(z_1), ..., f(z_k) $ are linearly dependent are $k$-degenerate, and G.a.f.'s for which
the random variables $ f(z_1), f'(z_1), ..., f(z_k), f'(z_k) $ are linearly dependent are
$2k$-degenerate.

Observe that the G.a.f.~\eqref{eq-10} satisfies condition~\eqref{eq-60} if and only
if $ Lf_n = 0 $ for every $ n $. So Gaussian Taylor series (infinite, or finite)
\[
f(z) = \sum_{n\ge 0} \zeta_n c_n z^n
\]
are $d$-nondegenerate provided that $c_0, c_1, ..., c_{d-1} \ne 0$.
Indeed, suppose that \eqref{eq-60} holds. Then
\[
Lz^n = \int_\gamma z^n r^L(z)\, {\rm d} z = 0, \qquad \text{for }n = 0, 1, ..., d-1\,.
\]
Choose $R$ so large that $G\subset \{|z|< R\}$. Then
\[
\int_{\{|z|=R\}} z^n r^L(z)\,
{\rm d} z = \int_\gamma z^n r^L(z)\, {\rm d} z = 0, \qquad \text{for }n = 0, 1, ..., d-1\,,
\]
whence,
\[
r^L(z) = O\bigl( z^{-(d+1)}\bigr), \qquad z\to\infty\,.
\]
That is, the degree of $r^L$ is not less than $d+1$, which means that $f(z)$ is $d$-nondegenerate,
as we claimed. In particular, we see that the G.E.F. is $d$-nondegenerate
for every non-negative integer $d$.

\begin{claim}\label{claim-B}
Let $f$ be a $d$-nondegenerate G.a.f., and let
$K\subset G$ be a compact set.
Then there exists positive constants $c(d, f, K)$ and $C(f, K)$ such that for each
functional $L$ of rank at most $d$ with $z_1, ..., z_k \in K$, we have
\begin{equation}\label{eq-70}
c \max_\gamma |r^L|^2 \le \E |Lf|^2 \le C \max_\gamma |r^L|^2\,.
\end{equation}
\end{claim}

\medskip\par\noindent{\em Proof of the upper bound in~\eqref{eq-70}:}
\begin{multline*}
\E | Lf |^2 = \E \left| \int_\gamma f r^L\, {\rm d}z \right|^2 \\
\le \left( \text{length}(\gamma) \cdot \max_\gamma |r^L| \right)^2 \cdot \int_\gamma \E |f|^2\, |{\rm d}z|
= C(f, \gamma) \left( \max_\gamma |r^L| \right)^2\,. \qquad \Box
\end{multline*}

\medskip The proof of the lower bound uses a simple compactness argument. Denote
by $R_d(K)$ the set of rational functions of degree at most $d$ vanishing at $\infty$ and having
all poles in $K$.

\begin{claim}\label{claim-C} Each sequence of rational functions $\{ r_m \}\subset R_d(K)$
with $\max_\gamma |r_m|~\le~1$ has a subsequence that converges uniformly on $\gamma$ to a function
$r\in R_d(K)$.
\end{claim}

\par\noindent{\em Proof of Claim~\ref{claim-C}:} We have $r_m = \tfrac{p_m}{q_m}$, where
$p_m$ and $q_m$ are polynomials of degree $\le d-1$ and $\le d$ correspondingly, and
$ q_m (z) = \prod_{i=1}^d (z-z_i(m)) $
with $z_1(m), ..., z_d(m)\!\in\!K$ for each $m$. Choosing a subsequence, we may assume that
$q_m$ converge uniformly on $\gamma$ to a polynomial $q$ of the same form. Then
$ \max_\gamma |p_m| \le C $ for each $m$. Since the degrees of the polynomials $p_m$ are
uniformly bounded, we can choose a subsequence that converges uniformly
on $\gamma$ to a polynomial of degree $\le d-1$.  \hfill $\Box$

\medskip\par\noindent{\em Proof of the lower bound in~\eqref{eq-70}:} Suppose that there
exists a sequence of functionals $L_m$ of the form~\eqref{eq-50} of rank $d$ or less,
with $\max_\gamma |r^{L_m}| = 1$ and with
$ \E |L_m f|^2 \to 0 $. By~\eqref{eq-45} and Claim~\ref{claim-C}, we can choose a subsequence of this
sequence such that $ r^{L_m} $ converge uniformly on $\gamma$ to some nonzero $r\in R_d(K)$
(to simplify our notation, we omit subindices).
Then $r=r^L$ for some functional $L$ of the same form~\eqref{eq-50}, and
\[
\lim_{m\to\infty} \E |L_m f - Lf|^2 \le
\left( \text{length}(\gamma) \right)^2 \cdot \int_\gamma \E |f|^2 |{\rm d} z| \cdot
\lim_{m\to\infty} \max_\gamma |r^{L_m}-r^L|^2 = 0\,.
\]
Hence, $ Lf = 0 $ almost surely, which contradicts the $d$-nondegeneracy of $f$.
\hfill $\Box$

\medskip
To estimate the $k$-point function, we use a
classical formula that goes back to Kac, Rice, and Hammersley,
see~\cite{BSZ} and~\cite[Chapter~3]{HKPV}:
\begin{equation}\label{eq-30}
\rho_f(z_1, ..., z_k) = \frac1{\pi^{2k} \det \Gamma_f}
\int_{\C^k} |\eta_1|^2 ... |\eta_k|^2  e^{-\frac12 \langle \Gamma_f^{-1} \eta', \eta' \rangle}
{\rm d}A(\eta_1) ... {\rm d} A(\eta_k),
\end{equation}
where $\Gamma_f = \Gamma_f(z_1, ..., z_k)$ is the covariance matrix of the random variables
\begin{equation}\label{eq-20}
f(z_1), f'(z_1), ..., f(z_k), f'(z_k)\,,
\end{equation}
and $\eta' = \left( 0, \eta_1, ..., 0, \eta_k \right)^{\tt T}$
is a vector in $\C^{2k}$.
Here, we assume that the random variables~\eqref{eq-20} are linearly independent; later, we will
impose on the G.a.f. $f(z)$ a somewhat stronger restriction of non-degeneracy. The
Kac-Rice-Hammersley formula~\eqref{eq-30} allows us to reduce estimates for the $k$-point function $\rho_f$
to estimates for the covariance matrix $\Gamma_f$.

We start with a simple observation. Consider ``special'' linear functionals of rank $2k$:
\begin{equation}\label{eq-40}
Lf = \sum_{j=1}^k \left[ \alpha_j f(z_j) + \beta_j f'(z_j) \right].
\end{equation}
\begin{claim}\label{claim-A} For every vector
$\delta = \left( \alpha_1, \beta_1, ... , \alpha_k, \beta_k \right)^{\tt T} $
in $\C^{2k}$, we have $ \langle \Gamma_f \delta, \delta \rangle = \E | Lf |^2 $.
\end{claim}

\par\noindent{\em Proof:} by straightforward inspection. \hfill $\Box$

\medskip Introduce the Gaussian polynomial
\[
f_{2k-1}(z) = \sum_{n=0}^{2k-1} \zeta_n z^n\,.
\]

\begin{claim}\label{claim-D}
Let $f$ be a $2k$-nondegenerate G.a.f.
and let $K\subset G$ be a compact set. Then there exists a positive constant $C(k, f, K)$
such that for each configuration
$z_1, ..., z_k\in K$,
\begin{equation}\label{eq-80}
C^{-1} \Gamma_{f_{2k-1}}(z_1, ..., z_k) \le \Gamma_f (z_1, ..., z_k) \le C \Gamma_{f_{2k-1}}(z_1, ..., z_k)\,,
\end{equation}
where the inequalities are understood in the operator sense.
\end{claim}

\par\noindent{\em Proof:} this is a straightforward consequence of
Claims~\ref{claim-A} and~\ref{claim-B}. \hfill $\Box$

\medskip Now, using the Kac-Rice-Hammersley formula~\eqref{eq-30}, we
readily get equivalence of the $k$-point functions $\rho_f$ and $\rho_{f_{2k-1}}$:

\begin{claim}\label{claim-E}
Let $f$ be a $2k$-nondegenerate G.a.f.
and let $K\subset G$ be a compact set. Then there exists a positive constant $C(k, f, K)$
such that for each configuration
$z_1, ..., z_k\in K$,
\begin{equation}\label{eq-90}
C^{-1} \rho_{f_{2k-1}}(z_1, ..., z_k) \le \rho_f (z_1, ..., z_k) \le C \rho_{f_{2k-1}}(z_1, ..., z_k)\,.
\end{equation}
\end{claim}

\par\noindent{\em Proof:}
First, note that Claim~\ref{claim-D} yields
\begin{equation}\label{eq-100}
C^{-2k} \det \Gamma_{f_{2k-1}} \le \det \Gamma_f \le C^{2k} \det \Gamma_{f_{2k-1}}\,.
\end{equation}
Then, plugging estimates~\eqref{eq-80} and~\eqref{eq-100} into~\eqref{eq-30}, we get
\begin{multline*}
\rho_f(z_1, ..., z_k) \le \frac{C^{2k}}{\pi^{2k} \det \Gamma_{f_{2k-1}}}
\int_{\C^k} |\eta_1|^2\,...\,|\eta_k|^2\, e^{-\frac12 C^{-1} \langle \Gamma_{f_{2k-1}}^{-1} \eta', \eta' \rangle}
\, {\rm d} A(\eta_1)\,...\, {\rm d} A(\eta_k) \\
\qquad \qquad \quad = \frac{C^{2k}}{\pi^{2k} \det \Gamma_{f_{2k-1}}}
\cdot C^{2k}
\int_{\C^k} |\eta_1|^2\,...\,|\eta_k|^2\, e^{-\frac12 \langle \Gamma_{f_{2k-1}}^{-1} \eta', \eta' \rangle}
\, {\rm d} A(\eta_1)\,...\, {\rm d} A(\eta_k) \\
= C^{4k} \rho_{f_{2k-1}}(z_1, ..., z_k)\,.
\end{multline*}
Similarly, we get the lower bound in \eqref{eq-90}. \hfill $\Box$

\medskip
We see that it suffices to estimate the $k$-point function
only for one special Gaussian polynomial $f_{2k-1}$ of degree $2k-1$.
First we consider the probability
density  $\mathfrak{p}_{f_{2k-1}}(z_1, ..., z_{2k-1})$
of the joint distribution of all zeroes of $f_{2k-1}$. This means
that $ \mathfrak{p}_{f_{2k-1}} $ is a symmetric function such that,
for any symmetric function $ S $ of $2k-1$ complex variables,
the expected value of $ S(z_1, ..., z_{2k-1}) $ equals
\[
\int_{\C^k} S(z_1, ..., z_{2k-1})
\mathfrak{p}_{2k-1} (z_1, ..., z_{2k-1})\, {\rm d}A(z_1) ... {\rm d}A(z_{2k-1})\,.
\]
The density $\mathfrak{p}_{f_{2k-1}}$
can be computed using a classical formula for the Jacobian of the transformation of
zeroes of the polynomial into its Taylor coefficients~\cite{BBL, FH}:
\[
\mathfrak{p}_{f_{2k-1}}(z_1, ..., z_{2k-1}) = C_k \prod_{1\le i < j \le 2k-1} |z_i-z_j|^2
\cdot \bigl( \sum_{0 \le j \le 2k-1} | \sigma_j |^2\bigr)^{-2k}\,,
\]
where $\sigma_j$'s are the coefficients of the polynomial
\[
\prod_{1 \le i \le 2k-1} (z-z_i) = \sum_{0\le j \le 2k-1} \sigma_j z^j \qquad (\text{here,\ }
\sigma_{2k-1}=1).
\]
That is,
\[
\mathfrak{p}_{f_{2k-1}}(z_1, ..., z_{2k-1}) = \prod_{1\le i < j \le k} |z_i-z_j|^2 \cdot
H(z_1, ..., z_{2k-1})\,,
\]
where $H$ is a non-negative continuous function.
\begin{claim}\label{claim-upper_bound_H}
There exists a positive constant $C(k, K)$ such that for all $z_1, ..., z_k\!\in\!K$,
and all $z_{k+1}, ..., z_{2k-1}\in \C$,
we have
\[
H(z_1, ..., z_{2k-1}) \le C(k, K) \prod_{i=k+1}^{2k-1} \left( 1 + |z_i| \right)^{-4}\,.
\]
\end{claim}

\medskip\par\noindent{\em Proof of Claim~\ref{claim-upper_bound_H}:} Let
\[ P(z)=\prod_{i=1}^{2k-1} (z-z_i)\,. \]
For each $z\in\T$, we have
\[
|P(z)| \le \sum_{j=0}^{2k-1} |\sigma_j|, \qquad {\rm whence} \quad
\sum_{j=0}^{2k-1} |\sigma_j|^2 \ge \frac1{2k} |P(z)|^2.
\]
An elementary geometry shows that
there exists a point $z\in\T$ so that $|z-z_i|\ge c_k$ for $i=1, ..., 2k-1$,
whence, $|z-z_i|\ge c_k (1+|z_i|)$ for $i=1, ..., 2k-1$, and
\[
|P(z)|^2 \ge c_k \prod_{i=1}^{2k-1} (1 + |z_i|)^2 \ge c_k \prod_{i=k+1}^{2k-1} (1 + |z_i|)^2\,.
\]
Therefore,
\[
\sum_{j=0}^{2k-1} |\sigma_j|^2 \ge c_k \prod_{i=k+1}^{2k-1} (1 + |z_i|)^2\,,
\]
and
\[
\mathfrak{p}_{f_{2k-1}} (z_1, ..., z_{2k-1}) \le C_k \prod_{1\le i < j \le 2k-1} |z_i-z_j|^2 \,
\, \cdot\, \prod_{i=k+1}^{2k-1} (1 + |z_i|)^{-4k}\,.
\]
At last, using that $|z_i-z_j|\le (1+|z_i|)(1+|z_j|)$, we get
\begin{multline*}
H(z_1, ..., z_{2k-1}) \le C(k, K) \prod_{i=k+1}^{2k-1} (1+|z_i|)^{2(2k-2)}\,
\prod_{i=k+1}^{2k-1} (1 + |z_i|)^{-4k} \\
= C(k, K) \prod_{i=k+1}^{2k-1} \left( 1 + |z_i| \right)^{-4}
\end{multline*}
completing the proof of the claim. \hfill $\Box$

\medskip Now, we readily finish the proof of the local universality theorem.
The $k$-point function $\rho_{f_{2k-1}} (z_1, ...., z_k)$
can be obtained from the probability density function $\mathfrak{p}_{f_{2k-1}}$
by integrating out the extra variables
\begin{multline*}
\rho_{f_{2k-1}}(z_1, ..., z_k) = C_k
\int_{\C^{k-1}} \mathfrak{p}_{f_{2k-1}}(z_1,..., z_{2k-1})\,
{\rm d} A(z_{k+1})\,...\, {\rm d} A(z_{2k-1}) \\
= C_k \prod_{1\le i < j \le k} |z_i-z_j|^2 \cdot
\int_{\C^{k-1}} H(z_1, ..., z_{2k-1})\,
{\rm d} A(z_{k+1})\,...\, {\rm d} A(z_{2k-1})\,.
\end{multline*}

Due to Claim~\ref{claim-upper_bound_H},
the integral on the right-hand side converges uniformly with respect to the variables
$z_1, ..., z_k\in K$.
Hence, it is a continuous function in $z_1$, ..., $z_k$.
Given $z_1$, ..., $z_k$, the function $H$ (as a function of variables $z_{k+1}$, ..., $z_{2k-1}$)
vanishes only on a set of zero Lebesgue measure in $\C^{k-1}$, hence, the integral is a positive
function of $z_1$, ..., $z_k$. This completes the proof of Theorem~\ref{thm-weak-univ}.
\hfill $\Box$

\section{Clustering of $k$-point functions. Proof of Theorem~\ref{thm_main*}}

The proof again employs the linear functionals $L$ introduced in \eqref{eq-40}
and Kac-Rice-Hammersley's formula~\eqref{eq-30}.
By $\rho\T$ we denote the circumference $\{ |z|=\rho \}$.

Let $F$ be a G.E.F.\,. For $w\in\C$, denote by $T_w$ the projective shift
\[ T_w F(z) = F(w+z) e^{-z\overline{w}} e^{-\frac12 |w|^2}\,. \]
Comparing the covariances $\E \left\{ F(z_1) \overline{F(z_2)} \right\} $
and $\E \left\{ T_w F(z_1) \overline{T_w F(z_2)} \right\} $, one readily checks
that for every $w\in\C$, $T_w F$ is also a G.E.F.\,.
Our first aim is to estimate the covariance
$ \E \left\{ L_1 \left( T_{w_1}F \right) \overline{ L_2 \left( T_{w_2} F \right) } \right\} $,
where $L_1$, $L_2$ are two linear functionals of the kind \eqref{eq-40}, and $w_1, w_2 \in \C$,
$w_1 \ne w_2$.

\begin{claim}\label{claim-intermed}
Let $k\ge 1$, $\rho>0$, $ |w_1-w_2| \ge 4\rho$,
and let the points $z_j$ participating in the definition of
the functional $L$ satisfy $|z_j|\le \rho$ ($j=1, ..., k$). Then
\begin{multline}\label{eq-*}
\left| \E \left\{ L_1 \left( T_{w_1}F \right) \overline{ L_2 \left( T_{w_2} F \right) } \right\} \right|
\\
\le C(\rho, k) e^{-\frac12 (|w_1-w_2| - 4\rho)^2}
\, \left( \E |L_1 T_{w_1} F|^2 + \E |L_2T_{w_2} F|^2 \right)\,.
\end{multline}
\end{claim}

Note that, since $T_w F$ is equidistributed with $F$ for all $w\in\C$, we
can also rewrite the sum on the right-hand side as $\E |L_1 F|^2 + \E |L_2 F|^2$.

\medskip\par\noindent{\em Proof of Claim~\ref{claim-intermed}}: Using the representation~\eqref{eq-45}
with $\gamma=2\rho\T$, we can write
\begin{multline*}
\left| \E \left\{ L_1 (T_{w_1}F) \overline{L_2(T_{w_2}F)} \right\} \right| \\
= \left| \iint_{2\rho\T \times 2\rho\T} r^{L_1}(z_1) \overline{r^{L_2}(z_2)}\,
\E \left\{ T_{w_1} F(z_1) \overline{T_{w_2} F(z_2)} \right\}\,  {\rm d}z_1\, {\rm d}z_2 \right|
\qquad \qquad \\
\le \frac12 (2\pi\rho)^2 \cdot \bigl( \max_{2\rho\T} |r^{L_1}|^2 + \max_{2\rho\T} |r^{L_1}|^2 \bigr)
\cdot \sup_{z_1, z_2\in 2\rho\T}
\left| \E \left\{ T_{w_1} F(z_1) \overline{T_{w_2} F(z_2)} \right\} \right|\,.
\end{multline*}
Since for any $w_1, w_2, \la_1, \la_2\in\C$, we have
\[
\left| \E \left\{ T_{w_1} F(\la_1) \overline{T_{w_2} F(\la_2)} \right\} \right|
= e^{\frac12 |\la_1|^2 + \frac12 |\la_2|^2 - \frac12 |(w_1+\la_1)-(w_2+\la_2)|^2}\,,
\]
the last supremum does not exceed
$ e^{4\rho^2} e^{-\frac12 (|w_1-w_2|-4\rho)^2}$,
provided that $|w_1-w_2|\ge 4\rho$. Taking into account that
$ | r^{L_i} |^2 \le C(k, \rho) \E | L_i F |^2 $ everywhere on $2\rho\T$ (Claim~\ref{claim-B}),
we get~\eqref{eq-*}. \hfill $\Box$

\medskip In what follows, we also need the following simple geometric claim.

\begin{claim}\label{claim-geom}
Suppose that $z_1, ..., z_k\in \C $, and
$\underline{d}\colon (0, +\infty) \to (0, +\infty)$ is any increasing
function. Then there exists $\rho \ge 1$  and a covering of points $z_j$ ($j=1, ..., k$)
by disks $D(w_n, \rho)$ ($n=1, ..., N$) such that $N\le k$ and $ |w_{n'}-w_{n''}|\ge
\underline{d}(\rho)$ for all $n'\ne n''$.
Moreover, $\rho$ is bounded from above by some constant depending on $\underline{d}$
and $k$ only.
\end{claim}

\medskip\par\noindent{\em Proof of Claim~\ref{claim-geom}:}
Take $\rho_1=1$ and consider the covering by the disks $D(w_n, \rho_1)$ with $N=k$, $w_n=z_n$
for all $n$. If all pairwise distances $|w_{n'}-w_{n''}|\ge \underline{d}(\rho_1)$, we are done.
If not, replace two centers $w_{n'}$ and $w_{n''}$ satisfying $|w_{n'}-w_{n''}| < \underline{d}(\rho_1)$
by one center $ \tfrac12 (w_{n'}+ w_{n''}) $ and increase all radii to
$\rho_2 = \rho_1+\frac12 \underline{d}(\rho_1)$. If in the resulting configuration all pairwise distances
between centers are at least $\underline{d}(\rho_2)$, we are done. Otherwise, again, replace two
close centers by one and increase the radii to $\rho_3 = \rho_2 + \frac12 \underline{d}(\rho_2)$, and so on.

The process will stop after at most $k-1$ steps, and the final radius will be bounded by the
$k$-th term in the recursive sequence given by
$ \rho_1=1$, and $\rho_{j+1} = \rho_j+\tfrac12 \underline{d}(\rho_j)$.
\hfill $\Box$

\begin{claim}\label{claim-key}
There exists a constant $\Delta=\Delta(k)$ with the following property. Let $z_1, ..., z_k\in\C$.
Suppose that the set of indices $\left\{ 1, 2, ..., k \right\}$ is partitioned into two
non-empty subsets $I$ and $J$ so that
\[ d = d(Z_I, Z_J) = \inf \left\{ |z_i-z_j|\colon i\in I, j\in J \right\} \ge \Delta\,.\]
Then for any two linear functionals
\[
L^I f = \sum_{i\in I} \bigl[ \alpha_i f (z_i) + \beta_i f'(z_i) \bigr], \quad
L^J f = \sum_{j\in J} \bigl[ \alpha_j f (z_j) + \beta_j f'(z_j) \bigr],
\]
one has
\begin{equation}\label{eq-cov}
\left| \E \left\{ \bigl( L^I F \bigr) \bigl( \overline{L^J F} \bigr) \right\} \right|
\le C(k) e^{-\frac12 (d-\Delta)^2}
\left\{ \E | L^I F|^2 + \E |L^J F |^2\right\}\,.
\end{equation}
\end{claim}

\medskip
\par\noindent{\em Proof of Claim~\ref{claim-key}:} We put
\[
\underline{d}(\rho) = 4\rho + \sqrt{2\log \left(10k\, C(\rho, k)\right) }
\]
where the constant $C(\rho, k)$ is the same as in \eqref{eq-*},
and apply the construction of Claim~\ref{claim-geom} to the points
$z_j$. We get a number $\rho \le \rho (k)$ with $\rho (k)\ge 1$ independent of $z_1, ..., z_k$,
and a sequence of disjoint disks $D(w_n, \rho)$ ($n=1, ..., N$, $N\le k$)
such that each point $z_j$ lies in at least one of these disks.
Then we take $\Delta = 6\rho (k)$, and notice that
no disk can contain two points $z_i$, $z_j$ with $i\in I$, and $j\in J$.
Thus, the set of indices $\left\{ 1, 2, ..., N \right\}$ gets partitioned into two non-empty
subsets $\mathcal I$ and $\mathcal J$ such that the disks corresponding to indices from $\mathcal I$ cover
the points $z_i$ with indices $i\in I$, and similarly for $J$ and $\mathcal J$.

Now, we put
\[
L_n f = \sum_{i\colon z_i\in D(w_n, \rho)} \bigl[ \alpha_i  f (z_i) + \beta_i f'(z_i) \bigr]
\]
and note that
$ L_n f = \left( L_n \circ T_{w_n}^{-1} \right) \left( T_{w_n} f \right)
= \widetilde{L}_n \left( T_{w_n} f \right)$,
where $\widetilde{L}_n$ is some linear functional of the kind~\eqref{eq-40}
in whose definition the points $z_i - w_n$ with $z_i\in D(w_n, \rho)$ participate
\footnote{More explicitly, if
$ L f = \sum_i \bigl[ \alpha_i  f (z_i) + \beta_i f'(z_i) \bigr] $,
then
\[
\widetilde{L}_n\, g =
\sum_{i\colon z_i\in D(w_n, \rho)}
\bigl[ (\alpha_i + \beta_i \overline{w}_n) e^{z_i \overline{w}_n}
e^{-\frac12 |w|^2}  g(z_i-w_n)
+ \beta_i e^{z_i \overline{w}_n} e^{-\frac12 |w|^2} g'(z_i-w_n)
\bigr]\,.
\]
}.
Observe that
\begin{multline*}
\E \left| L^I F \right|^2
= \sum_{n\in\mathcal I} \E \left| \widetilde{L}_n\, T_{w_n} F \right|^2
+ \sum_{n', n''\in\mathcal I,\, n'\ne n''}\ \underbrace{\E \left\{ \bigl( \widetilde{L}_{n'}\, T_{w_{n'}} F \bigr)\,
\bigl( \overline{\widetilde{L}_{n''}\, T_{w_{n''}} F} \bigr) \right\} }_{= \sigma (n', n'')} \\
=
\sum_{n\in\mathcal I} \E \left| \widetilde{L}_n\, F \right|^2
+ \sum_{n', n''\in\mathcal I, \, n'\ne n''}\ \sigma (n', n'')\,.
\end{multline*}
By Claim~\ref{claim-intermed}, for $n'\ne n''$, we have
\begin{multline*}
\left| \sigma (n', n'') \right| \le C(\rho, k) e^{-\frac12 (\underline{d}(\rho) - 4\rho)^2}
\left[ \E \left| \widetilde{L}_{n'} F  \right|^2 + \E \left| \widetilde{L}_{n''} F  \right|^2 \right]
\\
\le \frac1{10k} \left[ \E \left| \widetilde{L}_{n'} F  \right|^2
+ \E \left| \widetilde{L}_{n''} F  \right|^2 \right]
\end{multline*}
by our choice of $\underline{d}(\rho)$, so
\[
\E \left| L^I F \right|^2 \ge \frac12 \sum_{n\in\mathcal I} \E \left| \widetilde{L}_n F \right|^2\,.
\]
A similar estimate holds for $L^J$, whence,
\[
\E \left| L^I F \right|^2  + \E \left| L^J F \right|^2 \ge
\frac12 \sum_{n} \E \left| \widetilde{L}_n F \right|^2\,.
\]

On the other hand,
\[
\left| \E \left\{ \bigl( L^I F \bigr) \bigl( \overline{L^J F} \bigr) \right\} \right|
\le \sum_{n'\in\mathcal I, \, n''\in\mathcal J}\, \left| \sigma (n', n'') \right|\,.
\]
But for $n'\in\mathcal I$, $n''\in\mathcal J$, we have
\[
\left| w_{n'}-w_{n''} \right| \ge \left| z_i - z_j \right| - 2\rho \ge d-2\rho\,,
\]
where $z_i$ is any point in $D(w_{n'}, \rho)$, and $z_j$ is any point in $D(w_{n''}, \rho)$.
Using Claim~\ref{claim-intermed} again, we get
\[
\left| \sigma (n', n'') \right| \le C(\rho, k) e^{-\frac12 (d-6\rho)^2}
\left[ \E \left| \widetilde{L}_{n'} F  \right|^2
+ \E \left| \widetilde{L}_{n''} F  \right|^2 \right]\,,
\]
and
\[
\sum_{n'\in\mathcal I, \, n''\in\mathcal J}\, \left| \sigma (n', n'') \right|
\le C(\rho, k) e^{-\frac12 (d-6\rho)^2} \sum_n \E \left| \widetilde{L}_n F \right|^2\,.
\]
Recalling that $\rho\le \rho(k)$, we get the claim. \hfill $\Box$

\medskip Now, we are ready to prove Theorem~\ref{thm_main*}, that is,
the multiplicative clustering. To simplify our notation, we denote by
\[
\epsilon = \epsilon (d, k) = C(k) e^{-\frac12 (d-\Delta)^2}
\]
the small factor on the right-hand side of \eqref{eq-cov}. Recall that
$d$ is a shortcut for $d(Z_I, Z_J)$ and that $d\ge 2\Delta$.
In what follows, we assume that $\Delta$ is so large that $\epsilon < \tfrac12$.
We will show that
\begin{equation}\label{eq-new*}
\left[ \frac{1-\epsilon}{1+\epsilon} \right]^{2k} \rho(Z_I)\rho(Z_J)
\le \rho (Z)  \le
\left[ \frac{1+\epsilon}{1-\epsilon} \right]^{2k} \rho(Z_I)\rho(Z_J)\,,
\end{equation}
which yields Theorem~\ref{thm_main*}.

To prove~\eqref{eq-new*}, we repeat the argument already used in the proof of
Theorem~\ref{thm-weak-univ}, and interpret inequality~\eqref{eq-cov} from the
previous claim as a two-sided
estimate for the covariance matrix $\Gamma = \Gamma_F(z_1, ..., z_k)$
of the family of $2k$ Gaussian random variables $\left( F(z_i), F'(z_i) \right)$ ($1\le i \le k$).
Consider also the covariance matrices $\Gamma_I$ and $\Gamma_J$ of the subfamilies with $i\in I$
and $i\in J$ respectively. Then estimate~\eqref{eq-cov} can be rewritten in
the following form
\begin{equation}\label{eq-7}
(1-\epsilon) \Gamma_{I, J} \le  \Gamma \le (1+\epsilon) \Gamma_{I, J}\,,
\end{equation}
where
\[
\Gamma_{I, J} = \left(
\begin{matrix}
\Gamma_I & 0 \\
0 & \Gamma_J
\end{matrix}
\right)
\]
is the covariance matrix corresponding to two independent copies $F_I$ and $F_J$ of the
G.E.F. $F$. Inequalities~\eqref{eq-7} yield
\[
\left( 1-\epsilon \right)^{2k} \det \Gamma _{I, J}
\le \det \Gamma \le \left( 1+\epsilon \right)^{2k} \det \Gamma _{I, J}\,.
\]

We again use the Kac-Rice-Hammersley formula~\eqref{eq-30}:
\[
\rho (z_1, ..., z_k) = \frac1{\pi^{2k} \det\Gamma}
\int_{\C^k} |\eta_1|^2\,...\,|\eta_k|^2\, e^{-\frac12 \langle \Gamma^{-1} \eta', \eta' \rangle}
\, {\rm d} A(\eta_1)\,...\, {\rm d} A(\eta_k)\,,
\]
where $\eta' = \left( 0,  \eta_1, ..., 0, \eta_k \right)^{\tt T} $ is a vector in
$\C^{2k}$. Note now that
\[
\left( 1+\epsilon \right)^{-2k} \left( \det \Gamma_I \right)^{-1}
\left( \det \Gamma_J \right)^{-1}
\le \left( \det \Gamma \right)^{-1} \le
\left( 1-\epsilon \right)^{-2k} \left( \det \Gamma_I \right)^{-1}
\left( \det \Gamma_J \right)^{-1}\,,
\]
and that $\pi^{2k} = \pi^{2i} \cdot \pi^{2j}$, where $i= |I|$ and $j=|J|$.
To compare the integrals, observe that
\begin{multline*}
\int_{\C^k} |\eta_1|^2\,...\,|\eta_k|^2\, e^{-\frac12 \langle \Gamma^{-1} \eta', \eta' \rangle}
\, {\rm d} A(\eta_1)\,...\, {\rm d} A(\eta_k) \\
\ge \int_{\C^k} |\eta_1|^2\,...\,|\eta_k|^2\,
e^{-\frac12 (1-\epsilon)^{-1} \langle \Gamma_{I, J}^{-1} \eta', \eta' \rangle}
\, {\rm d} A(\eta_1)\,...\, {\rm d} A(\eta_k) \\
= \left( 1-\epsilon \right)^{2k} \int_{\C^k} |\eta_1|^2\,...\,|\eta_k|^2\,
e^{-\frac12 \langle \Gamma_{I, J}^{-1} \eta', \eta' \rangle}
\, {\rm d} A(\eta_1)\,...\, {\rm d} A(\eta_k) \\
=
\left( 1-\epsilon \right)^{2k} \int_{\C^{i}} |\eta_1|^2\,...\,|\eta_{i}|^2\,
e^{-\frac12 \langle \Gamma_I^{-1} \eta_I', \eta_I' \rangle}
\, {\rm d} A(\eta_1)\,...\, {\rm d} A(\eta_{i}) \\
\times
\int_{\C^{j}} |\eta_1|^2\,...\,|\eta_{j}|^2\,
e^{-\frac12 \langle \Gamma_J^{-1} \eta_J', \eta_J' \rangle}
\, {\rm d} A(\eta_1)\,...\, {\rm d} A(\eta_{j})\,,
\end{multline*}
where
$\eta' = \left( 0,  \eta_1, ..., 0, \eta_k \right)^{\tt T} $,
$\eta_I' = \left( 0,  \eta_1, ..., 0, \eta_i \right)^{\tt T} $, and
$\eta_J' = \left( 0,  \eta_1, ..., 0, \eta_j \right)^{\tt T} $.
Similarly, we get an upper bound of the same kind with the
factor $(1+\epsilon)^{2k}$. This proves~\eqref{eq-new*} and hence
proves Theorem~\ref{thm_main*}. \hfill $\Box$

\section{Central limit theorem via clustering and cumulants. \\
Proof of Theorem~\ref{thm_norm}}

The proof of Theorem~\ref{thm_norm} is based on the estimates of cumulants $s_k = s_k(h)$ of the
linear statistics $ n(h) = \sum_{a\in\mathcal Z} h(a)$
defined by the formal expansion
\[ \log\, \E \bigl\{ e^{t n(h)} \bigr\} = \sum_{k\ge 1} \frac{s_k (h)}{k!} t^k\,. \]
We will show that for each bounded measurable function $h$ with compact support $ {\rm spt}(h) $
and for each $k\ge 2$,
\[
|s_k(h)| \le C_k A({\rm spt}(h)) \| h \|_\infty^k\,.
\]
The key point of this estimate is that applying it to the scaled statistics $ n(R; h) $ we get
the factor $R^2$ on the right-hand side. Together with the growth of the standard
deviation $ \sigma (R; h) $ as a positive power of $R$, this will yield the asymptotic normality
of $ n(R; h) $ when $ R\to\infty $.

\medskip
First, we introduce some notation and recall well-known facts about the moments and cumulants.
We denote by $\Pi(k, j)$ the collection of all unordered partitions of the set $\{1, 2, ..., k\}$
into $j$ disjoint nonempty blocks, and by $\Pi(k)$ the collection of all unordered partitions of
the set $\{1, 2, ..., k\}$ into disjoint nonempty blocks. For $\pi\in\Pi(k)$, we denote the
blocks by $\{\pi_1, ..., \pi_j\}$ with an arbitrarily chosen enumeration,
and denote the lengths of the blocks by $p_t=|\pi_t|$,
$1\le t\le j$. The number of blocks in the partition $\pi$ will be denoted by $|\pi|$.

The moments $m_k$ and the cumulants $s_k$ are related to each other by classical
formulas~\cite[Chapter~II, \S~12]{Shiryaev}:
\begin{equation}\label{eq_mc}
s_k = \sum_{\ell=1}^k (-1)^{\ell-1} (\ell-1)! \sum_{\pi\in\Pi(k, \ell)} m_{p_1}\, ...\, m_{p_\ell}\,.
\end{equation}
The inverse formula has the form
\begin{equation}\label{eq_cm}
m_k = \sum_{\pi\in\Pi(k)} s_{p_1} \, ...\, s_{p_t}\,.
\end{equation}

Next, we recall the definition of so called {\em truncated} (or Ursell)
$k$-point functions $\rho_k^T$. The truncated $k$-point functions play the same r\^{o}le
for the cumulants as the usual $k$-point functions for the moments.
Their definition is suggested by~\eqref{eq_mc}. They are symmetric functions
\[
\rho_k^T\colon \bigl\{Z\!=\!(z_1, ..., z_k)\!\in\!\C^k\colon z_i\ne z_j, {\ \rm for\ } i\ne j  \bigr\} \to \R
\]
defined by the formula
\begin{equation}\label{eq_mc1}
\rho^T_k (Z) =
\sum_{\ell=1}^k (-1)^{\ell-1} (\ell-1)! \sum_{\pi\in\Pi(k, \ell)} \rho_{p_1} (Z_{\pi_1})\, ...\,
\rho_{p_\ell} (Z_{\pi_\ell})\,.
\end{equation}

In particular, $\rho_1^T = \rho_1$,
\begin{equation}\label{eq-2trunc}
\rho_2^T(z_1, z_2) = \rho_2(z_1, z_2) - \rho_1 (z_1) \rho_1(z_2)\,,
\end{equation}
then
\begin{multline*}
\rho_3^T(z_1, z_2, z_3) = \rho_3(z_1, z_2, z_3) \\
- \bigl(\, \rho_2(z_1, z_2) \rho_1(z_3) +
\rho_2(z_1, z_3) \rho_1(z_2) + \rho_2(z_2, z_3) \rho_1(z_1)\,
\bigr) \\
+ 2 \rho_1 (z_1) \rho_1(z_2) \rho_1(z_3)\,,
\end{multline*}
and so on\, ...\,. The inversions to \eqref{eq_mc1} look the same as in \eqref{eq_cm}:
\begin{equation}\label{eq_cm1}
\rho_k (Z)  = \sum_{\pi\in\Pi(k)} \rho^T_{p_1}(Z_{\pi_1}) \, ...\, \rho^T_{p_t}(Z_{\pi_t})\,.
\end{equation}
See e.g. \cite[Appendix~A.7]{Mehta} for the proof of the equivalence of the formulas~\eqref{eq_mc1}
and \eqref{eq_cm1} based on generating functions; another short and direct proof of
this equivalence can be found in \cite[\S~2]{BH}.

\medskip
Asymptotic factorization of $k$-point correlation functions expressed in \eqref{eq_def_cluster}
yields fast asymptotic decay of truncated $k$-point functions when the diameter of
the configuration gets large.
\begin{claim}\label{claim-cor1}
Suppose that the $k$-point functions $\rho_k$ of a point process
$\mathcal Z$ are clustering, i.e., satisfy~\eqref{eq_def_cluster} with a fast decreasing
function $\phi$, and are bounded. Then,
\begin{equation}\label{eq-trunc_estimate0}
\left| \rho_k^T (Z) \right| \le C_k \widetilde{\phi}\!\left( c_k {\rm diam}(Z) \right),
\end{equation}
where $ \widetilde{\phi}\!=\!\min(1, \phi) $ and
${\rm diam}(Z)\!=\!\displaystyle\max_{1\le i<j\le k} |z_i-z_j|$ is the diameter
of the configuration~$Z$.
\end{claim}

\medskip\par\noindent{\em Proof of Claim~\ref{claim-cor1}}: We
show that for each $k\ge 2$ and for each partition of the set of indices
$\{1, 2, ..., k\} = I \sqcup J$,
\begin{equation}\label{eq-trunc_estimate1}
\left| \rho_k^T (Z) \right| \le C_k \widetilde{\phi}\!\left( c_k d(Z_I, Z_J)\right).
\end{equation}
Since for each configuration $Z$, there exists a partition $\{1, 2, ..., k\} = I \sqcup J$
with $d(Z_I, Z_J) \ge c_k {\rm diam}(Z)$, estimate~\eqref{eq-trunc_estimate1}
yields~\eqref{eq-trunc_estimate0}.

To prove \eqref{eq-trunc_estimate1}, we
use induction on $k$. For $k=2$, estimate \eqref{eq-trunc_estimate1}
follows from equation~\eqref{eq-2trunc}, clustering of $\rho_2$, and
boundedness of $\rho_1$ and $\rho_2$.
Now, suppose that $k\ge 3$. We fix a partition $\{1, 2, ..., k\} = I \sqcup J$ with $i=|I|$ and
$j=|J|$. We say that a partition $\pi\in\Pi(k)$
{\em refines} the partition $I \sqcup J$ if each set $\pi_s$ is a subset of either $I$, or $J$.
Otherwise, we say that the partition $\pi$ {\em mixes} $I \sqcup J$.
By the inversion formula~\eqref{eq_cm1}, we have
\[
\rho_k^T (Z) = \rho_k (Z) -
\sum_{t=|\pi|\ge 2} \rho_{p_1}^T (Z_{\pi_1})\, ...\, \rho_{p_t}^T (Z_{\pi_t}),
\]
and
\[
\rho_{i} (Z_I) \rho_{j} (Z_J) =
\sum_{\pi {\rm\ refines\ } I \sqcup J, \  t=|\pi|\ge 2} \rho_{p_1}^T(Z_{\pi_1}) \, ...\,
\rho_{p_t}^T(Z_{\pi_t}).
\]
Therefore,
\[
\rho_k^T (Z)
= \rho_k(Z) - \rho_{i} (Z_I) \rho_{j} (Z_J) \, +
\sum_{\pi {\rm\ mixes\ } I \sqcup J, \  t=|\pi|\ge 2} \rho_{p_1}^T(Z_{\pi_1}) \, ...\,
\rho_{p_t}^T(Z_{\pi_t}).
\]
By the induction assumption, each term in the sum on the RHS
contains at least one factor which is bounded by the right-hand side of~\eqref{eq-trunc_estimate1},
while the other factors are bounded by constants.
\hfill $\Box$

\begin{claim}\label{claim-cor2}
Suppose that the truncated $k$-point functions $\rho_k^T$ of a point process
$\mathcal Z$ satisfy~\eqref{eq-trunc_estimate0} with a fast decreasing
function $\phi$. Then, for each $k\ge 2$,
\begin{equation}\label{eq-trunc_estimate2}
\sup_{z_k\in\C}\, \int_{\C^{k-1}} \left| \rho_k^T (Z) \right|\,
{\rm d}A(z_1) ... {\rm d}A(z_{k-1}) < \infty\,.
\end{equation}
\end{claim}

\par\noindent{\em Proof of Claim~\ref{claim-cor2}:} We fix a point $z_k\in \C$,
and set $Z=(Z', z_k)$, where $Z'\in\C^{k-1}$. Then we
split $\C^{k-1}$ into disjoint sets $G_\ell$:
\[
G_0 = \left\{Z'\colon {\rm diam}(Z) \le 1 \right\}, \ {\rm and\ }
G_\ell = \left\{Z'\colon 2^{\ell-1} < {\rm diam}(Z) \le 2^\ell \right\} \ {\rm for\ } \ell\ge 1.
\]
Applying estimate~\eqref{eq-trunc_estimate0}, we get
$ \bigl| \rho_k^T (Z) \bigr| \le C_k \phi \left( c_k 2^{\ell-1} \right) $
provided that $Z'\!\in\!G_\ell$ with $\ell\!\ge\!1$. Then the integral of
the absolute value of $ \rho_k^T $ over $G_\ell$ does not exceed
\[ C_k 2^{2\ell (k-1)} \widetilde{\phi} \bigl( c_k 2^{\ell-1} \bigr). \]
Hence, the integral on the left-hand side of~\eqref{eq-trunc_estimate2} does not exceed
\[
C_k + C_k \sum_{\ell\ge 1} 2^{2\ell (k-1)} \widetilde{\phi} \bigl( c_k 2^{\ell-1} \bigr) \lesssim C_k
\]
(in the last estimate we use that the function $\phi$ decreases faster than any power). \hfill $\Box$

\medskip
The computation of the $k$-th cumulant $ s_k(h) $
of the linear statistics $ n(h) $ requires knowledge of $j$-point truncated functions with all $j\le k$.
For bounded compactly supported measurable functions $h_1$, ..., $h_k$ on $\C$,
we put
\[
\langle h_1, ..., h_k
\rangle^T \stackrel{\rm def}= \int_{\C^k} h_1(z_1)\,...\,h_k(z_k)\, \rho^T (z_1, ..., z_k)\,
{\rm d}A(z_1)...{\rm d}A(z_k)\,.
\]
Given a set $P=\{ p_1, ..., p_k \}$ of $k$ positive integers, we define
\[
s(h, P) = \langle h^{p_1}, ..., h^{p_k}
\rangle^T\,.
\]
Given a partition $\gamma$, we denote by $P_\gamma = \{|\gamma_1|, ..., |\gamma_j|\}$
the set whose elements are the lengths of the blocks of $\gamma$.
\begin{claim}\label{claim-moments-cumulants}
We have
\begin{equation}\label{eq-s_k}
s_k(h) =
\sum_{\gamma\in\Pi(k)}\, s(h, P_\gamma)\,.
\end{equation}
\end{claim}

Probably, these relations are well known to those who worked with $k$-point functions
of point processes. For this reason, we relegate the proof of Claim~\ref{claim-moments-cumulants}
to the appendix.

Now, we easily estimate the cumulants of the linear statistics $ n(h)$:
\begin{claim}\label{claim-tr_estimate}
For each bounded compactly supported function $h$, we have
\[
|s_k(h)| \le C_k \, \| h \|_\infty^k \, A({\text spt}(h))\,.
\]
\end{claim}

\medskip\par\noindent{\em Proof of Claim~\ref{claim-tr_estimate}}:
By equation~\eqref{eq-s_k}, it suffices to estimate from above
$ \left| \langle h^{p_1}, ..., h^{p_j} \rangle^T \right| $
where $p_1,  ..., p_j$ are positive integers such that
$\sum_j p_j = k$.

Let $G=\text{spt}(h)$. Then
\begin{multline*}
\left| \langle h^{p_1}, ..., h^{p_j} \rangle^T \right|
\le \int_{\C^j} | h (z_1) |^{p_1} \,...\, | h (z_j) |^{p_j} \, |\rho^T (z_1, ..., z_j) |\,
{\rm d}A(z_1)...{\rm d}A(z_j) \\
\le \| h \|_\infty^k \, \int_{\C^j} \done_G (z_1) \, ...\, \done_G (z_j) \,
|\rho^T (z_1, ..., z_j) |\, {\rm d}A(z_1)...{\rm d}A(z_j) \\
\qquad\qquad\qquad
\le \| h \|_\infty^k \int_\C \done_G(z_j) \, {\rm d}A(z_j)\, \int_{\C^{j-1}} |\rho^T (z_1,\, ...\, z_j)|\,
{\rm d}A(z_1) ... {\rm d}A(z_{j-1}) \\
\stackrel{\eqref{eq-trunc_estimate2}}\le C_k \| h \|_\infty^k\, \int_{\C} \done_G (z_j) {\rm d} A(z_j)
= C_k \| h \|_\infty^k\, A(G)\,,
\end{multline*}
proving the claim. \hfill $\Box$

\medskip\par\noindent{\em Proof of Theorem~\ref{thm_norm}}:
Denoting by $s_k(R; h)$ the cumulants of $n(R; h)$, and by
$s_k^*(R, h)$ the cumulants of the normalized random variables $n^* (R; h)$, we
have $s_1^*=0$, $s_2^*=1$, and for $k\ge 3$,
\[
s_k^*(R; h) = \frac{s_k(R; h)}{\sigma (R; h)^k}\,.
\]
By the estimate from the previous claim, we have
\[
|s_k(R; h)| \le C(k, h) R^2\,.
\]
Recalling that $\sigma (R; h)$ grows as a power of $R$, we see that for large enough $k$'s,
\[
\lim_{R\to\infty} s_k^*(R; h) = 0\,.
\]
By a version of the classical theorem of Marcinkiewicz (see \cite{Janson1, Soshn}),
this suffices to conclude that the random variables $ n^*(R; h) $ converge
in distribution to the Gaussian law when $R\to\infty$. This finishes off
the proof of Theorem~\ref{thm_norm}. \hfill $\Box$

\section{Lower bound for the variance.
Proof of Lemmas~\ref{lemma_var-growth} and \ref{lemma-simple}}

First, we prove Lemma~\ref{lemma-simple}.

\medskip\par\noindent{\em Proof of Lemma~\ref{lemma-simple}}:
It suffices to prove the lemma for $R=1$. The general case
readily follows by scaling
\[
x\mapsto R x, \quad h \mapsto h(R^{-1}\,\cdot), \quad \xi \mapsto R^{-1}\xi, \quad
\widehat{h} \mapsto R^2\, \widehat{h}(R\,\cdot)\,.
\]
Recall that $\rho (z_1, z_2) = r(z_1-z_2)$, and let
$\kappa (z) \stackrel{\rm def}= r(z)-1$. By the assumptions of the lemma, this is an $L^1(\R^2)$-function.
We have
\begin{multline*}
\sigma(1; h)^2 = \iint_{\R^2\times \R^2} h(z_1) h(z_2) \left[ \rho (z_1, z_2) + \delta (z_1-z_2) - 1\right]\,
{\rm d}A(z_1) {\rm d}A(z_2) \\
= \iint_{\R^2\times\R^2} h(z_1) h(z_2) \left[ \kappa (z_1-z_2) + \delta (z_1-z_2) \right]\,
{\rm d}A(z_1) {\rm d}A(z_2) \\
= \int_{\R^2} \bigl| \widehat{h}(\xi) \bigr|^2  \left[ 1 + \widehat{\kappa}(\xi)  \right]\,
{\rm d}A(\xi)
\end{multline*}
(in the last line we used Parseval's formula). By the Riemann-Lebesgue lemma, $\widehat{\kappa}(\xi)\to 0$
when $|\xi|\to\infty$. Thus, choosing a sufficiently big constant $C$ so that $|\widehat{\kappa}(\xi)|\le \tfrac12$
for $|\xi|\ge C$, we get
\[
\sigma(1; h)^2 \ge \frac12 \int_{|\xi|\ge C} \bigl| \widehat{h}(\xi) \bigr|^2  \, {\rm d}A(\xi)\,,
\]
proving the lemma. \hfill $\Box$

\begin{remark}[cf. Martin-Yal\c{c}in~\cite{MY}]
{\rm The computation above gives us
\begin{multline*}
\sigma(R; h)^2 = R^2\, \int_{\R^2} \bigl| \widehat{h}(\xi) \bigr|^2
\left[ 1 + \widehat{\kappa} ( R^{-1} \xi )  \right]\,
{\rm d}A(\xi) \\
= \left[ 1 + \widehat{\kappa}(0) +o(1) \right]\, \| h \|^2_{L^2(\R^2)}\, R^2\,,
\end{multline*}
as $ R\to\infty $. We conclude that translation-invariant point processes with $r-1\in L^1(\R^2)$
can be divided
into two groups: the processes with the Poissonian behaviour of the variance of linear statistics,
when $\widehat{\kappa}(0)\ne -1 $, i.e.,
\[
\int_{\R^2} \left[ r(x) - 1 \right]\, {\rm d}A(x) \ne -1\,,
\]
and the processes with non-Poissonian behaviour of the variance of linear
statistics, when $\widehat{\kappa}(0) = -1 $, i.e.,
\begin{equation}\label{eq-suppress_fluct}
\int_{\R^2} \left[ r(x) - 1 \right]\, {\rm d}A(x) = -1\,.
\end{equation}
Sometimes, the latter processes are called ``superhomogeneous processes''.
Lemma~\ref{lemma_var-growth} proven below is trivial for the processes from the first group.
On the other hand, though condition~\eqref{eq-suppress_fluct}
looks like a degeneration, many interesting translation-invariant
processes including our main hero, the zero point process $\mathcal Z_F$, and particle
processes based on Coulomb interaction that occur in statistical mechanics
belong to the second group.
}
\end{remark}

\begin{remark}
{\rm For the random zero process $\mathcal Z_F$,
the computation started above can be continued. This way, we arrive at an explicit formula
for the variance~\cite{NS1} which yields that
\[
\sigma(R; h)^2 \simeq R^{-2} \int_{|\xi|\le R} \bigl| \widehat{h}(\xi) \bigr|^2 |\xi|^4 \, {\rm d}A(\xi)
+ R^2 \int_{|\xi|\ge R} \bigl| \widehat{h}(\xi) \bigr|^2 \, {\rm d}A(\xi)\,.
\]
}
\end{remark}

Now, we prove the remaining Lemma~\ref{lemma_var-growth}.

\medskip\par\noindent{\em Proof of Lemma~\ref{lemma_var-growth}}:
Fix a multiplier
$m\in C_0^\infty (\R^2)$ with the following properties: $0\le m \le 1$ everywhere,
$m (\xi) = 0$ for $|\xi|\ge 2$, and $m(\xi) =1$ for $|\xi|\le 1$, and set
$m\ci R(\xi) = m( R^{-1}\xi )$.
We use the function $m_R$ to cut high-frequency oscillations in the spectrum of the indicator
function $\done\ci E$. Denote by $\phi\ci {E, R}$ the inverse Fourier transform of the product
$\widehat{\done\ci E} \cdot m\ci R$.
Since $\nabla \widehat{m\ci R}(x) = R^3 ( \nabla \widehat{m} )(Rx)$, we have
$ \| \nabla \widehat{m\ci R} \|_{L^1} = R \| \nabla \widehat{m} \|_{L^1}$, whence
\begin{equation}\label{eq_grad_phi}
| \nabla \phi\ci {E, R} | \lesssim R\,.
\end{equation}

By Lemma~\ref{lemma-simple} we have
\[
\sigma (R, \done\ci E)^2 \gtrsim R^2 \int_{|\xi|\ge R'} |\widehat{\done\ci E}|^2
\ge R^2 \int_{\R^2} | \widehat{\done\ci E}|^2 |1-m\ci {R'}|^2 = R^2 \int_{\R^2} |\done\ci E - \phi\ci {E, R'}|^2\,,
\]
with $R'=CR$, where $C$ is a constant on the right-hand side
of estimate~\eqref{eq_variance} in the statement of Lemma~\ref{lemma-simple}.
We estimate the integral on right-hand side twice. The first bound works when
$ A\left( \{\phi\ci {E, R'}\ge \tfrac12\} \right) \ge \tfrac12 A(E) $:
\begin{multline*}
\int_{\R^2} |\done\ci E - \phi\ci {E, R'}|^2
\ge \int_{\{\phi\ci {E, R'} \le \frac12 \}} |\done\ci E - \phi\ci {E, R'}|^2 \\
\ge \frac14 \int_{\{ \phi\ci {E, R'} \le \frac12 \}} 1
\stackrel{\eqref{eq_grad_phi}}\gtrsim
\frac1{R} \int_{\{ \phi\ci {E, R'} \le \frac12 \}} | \nabla \phi\ci {E, R'} | \qquad \qquad \\
\qquad \qquad \qquad \qquad
= \frac1{R }
\int_0^{1/2} {\rm Length}\left( \{ \phi\ci {E, R'} = t \} \right)\, {\rm d} t  \hfill ({\rm coarea\ formula}) \\
\qquad \qquad \qquad \qquad
\ge \frac1{R } \int_0^{1/2} \sqrt{ 4\pi A\left( \{ \phi\ci {E, R'} \ge t \} \right)}\, {\rm d} t \hfill
({\rm isoperimetric\ inequality}) \\
\gtrsim \frac1{R} \sqrt{ A\left( \{ \phi\ci {E, R'} \ge \tfrac12 \} \right)} \ge \frac1{R\sqrt 2}\, \sqrt{A(E)}\,.
\end{multline*}

On the other hand, if $  A\left( \{\phi\ci {E, R'}\ge \tfrac12\} \right) \le \tfrac12 A(E) $, we have
\begin{multline*}
\int_{\R^2} |\done\ci E - \phi\ci {E, R'}|^2 \ge \int_{\{\phi\ci {E, R'}
\le \frac12\}} |\done\ci E - \phi\ci {E, R'}|^2 \\
\ge \frac14 \int_{\{\phi\ci{E, R'} \le \frac12\}} \done\ci E
\gtrsim A(E) - A\left( \{\phi\ci {E, R'}\ge \tfrac12\} \right)  \ge \frac12 A(E)\,.
\end{multline*}
In both cases,
\[
R^2 \int_{|\xi|\ge 2R} |\widehat{\done\ci E}|^2 \gtrsim \min\left\{ A(E)R^2, \sqrt{A(E)}R  \right\}\,,
\]
proving the lemma.
\hfill $\Box$

\section*{Appendix: Proof of Claim~\ref{claim-moments-cumulants}}

First, we prove relations analogous to \eqref{eq-s_k}
that express the $k$-th moment $m_k(h)$ of the linear statistics
$n(h)$ in terms of $j$-point functions $\rho_j$ with $j\le k$.
For bounded compactly supported functions
$h_1$, ..., $h_k$ on $\C$, we put
\[
\langle h_1, ..., h_k \rangle \stackrel{\rm def}= \int_{\C^k} h_1(z_1)\,...\,h_k(z_k)\, \rho (z_1, ..., z_k)\,
{\rm d}A(z_1)...{\rm d}A(z_k).
\]
Given a set $P=\{ p_1, ..., p_k \}$ of $k$ positive integers, we define
\[
m(h, P) = \langle h^{p_1}, ..., h^{p_k} \rangle\,.
\]
As above, given a partition $\gamma$, we denote by $P_\gamma = \{|\gamma_1|, ..., |\gamma_j|\}$
the set whose elements are the lengths of the blocks of $\gamma$. Then
\begin{equation}\label{eq-m_k}
m_k(h) = \sum_{\gamma\in\Pi(k)}\, m(h, P_\gamma)\,.
\end{equation}

To prove this relation, we
denote by $( a_1, ..., a_k )$ an ordered
finite sequence of length $k$ of points in $\mathcal Z$,
possibly with repetitions.
Then
\[
m_k(h) = \E \left\{ \sum_{( a_1, ..., a_k )} h(a_1) ... h(a_k) \right\}
\]
where the sum on the right-hand side is taken over all finite sequences $( a_1, ..., a_k )$.
We fix a partition $\gamma\in\Pi (k)$ and say that
the sequence $( a_1, ..., a_k )$ is {\em subordinated} to $\gamma$ if the following
condition holds: {\em $a_s=a_t$ for $s\ne t$ if and only if the indices $s$ and $t$ belong to
the same block of the partition $\gamma$}. In this case, we'll write
$( a_1, ..., a_k )\lhd \gamma$. Clearly, each finite sequence of length $k$ is subordinated
to one and only one partition $\gamma\in\Pi (k)$.

Given a partition $\gamma\in\Pi(k, j)$, we fix an arbitrary enumeration
$\gamma_1, ..., \gamma_j$ of its blocks. Then we have
\[
\sum_{( a_1, ..., a_k )\lhd \gamma} h(a_1) ... h(a_k) =
\sum_{ \substack{ ( b_1, ..., b_j )\\ b_s \ne b_t \text{ for } s\ne t}} \
h^{|\gamma_1|}(b_1) ... h^{|\gamma_j|}(b_j),
\]
and then
\[
\E \left\{ \sum_{( a_1, ..., a_k )} h(a_1) ... h(a_k) \right\}
= \sum_{j=1}^k\, \sum_{\gamma\in\Pi(k, j)}\,
\E \left\{ \sum_{ \substack{(b_1, ..., b_j ) \\ b_s \ne b_t \text{ for } s\ne t}} \
h^{|\gamma_1|}(b_1) ... h^{|\gamma_j|}(b_j) \right\}\,.
\]
Since
\[
\E \left\{ \sum_{ \substack{ ( b_1, ..., b_j )\\ b_s \ne b_t \text{ for } s\ne t}} \
h_1(b_1) ... h_j(b_j) \right\} =
\bigl\langle h_1, ..., h_j \bigr\rangle
\]
(see for instance, \cite[Section~1.2]{HKPV}), we complete the proof of
relations~\eqref{eq-m_k}.
\hfill $\Box$

\medskip Now, we turn to the proof of relations~\eqref{eq-s_k}.
Given a partition $\gamma\in\Pi (k, j)$ we fix an arbitrary enumeration $\gamma_1$, ...,
$\gamma_j$ of its blocks. Then
according to the definition \eqref{eq_mc1} of truncated functions $\rho^T$, we have
\begin{multline}\label{eq-mult}
s(h, P_\gamma) = \sum_{\ell=1}^j (-1)^{\ell-1} (\ell-1)! \\
\times \sum_{\pi\in\Pi(j, \ell)} \int_{\C^j} h^{|\gamma_1|}(z_1) ... h^{|\gamma_j|}(z_j)\,
\rho (Z_{\pi_1}) ... \rho (Z_{\pi_\ell})\, {\rm d}A(z_1) ... {\rm d}A(z_j)
\end{multline}
where the partition $\pi$ splits the set of blocks
$ \{ \gamma_1, ..., \gamma_j \}  $ into $\ell$
`super-blocks' $\Pi^t$, $1\le t \le \ell$,
which we also enumerate arbitrarily. By $p_t$, $1\le t \le \ell$,
we denote the number of blocks of $\gamma$ that are included into the super-block $\Pi^t$.
By $ P^t = P^t_{\gamma, \pi} = \left\{ |\gamma_{\alpha_1}|, ..., |\gamma_{\alpha_{p_t}}| \right\} $
we denote the set of the lengths of the blocks $\gamma_{\alpha_1}$, ..., $\gamma_{\alpha_{p_t}}$
that are included into the super-block $\Pi^t$, this is the set of $p_t$ positive integers.
For instance, let
$k=8$, $j=3$, $\ell = 2$,
let
$\gamma_1 = \{ 1, 2, 3 \}$, $\gamma_2 = \{ 4, 5, 6 \}$, $\gamma_3 = \{ 7, 8 \}$,
and let
$ \pi_1 = \{ 1, 2 \}$, $\pi_2 = \{ 3 \}$.
Then
$ \Pi^1 = \{ \gamma_1, \gamma_2 \}$, $\Pi^2 = \{ \gamma_3 \}$,
$ p_1 = 2$, $p_2 = 1$, and $P^1 = \{3, 3\}$, $P^2 = \{2\}$.

\medskip
We factor the integrals on the right-hand side of~\eqref{eq-mult}
\[
\int_{\C^j} h^{|\gamma_1|}(z_1) ... h^{|\gamma_j|}(z_j)\,
\rho (Z_{\pi_1}) ... \rho (Z_{\pi_\ell})\, {\rm d}A(z_1) ... {\rm d}A(z_j)
= \prod_{t=1}^\ell m\left( h, P^t_{\gamma, \pi} \right).
\]
Then the right-hand side of~\eqref{eq-s_k} equals
\begin{equation}\label{eq-long}
\sum_{j=1}^k\, \sum_{\gamma\in\Pi(k, j)}\,
\sum_{\ell=1}^j (-1)^{\ell-1} (\ell-1)!
\sum_{\pi\in\Pi(j, \ell)}\,
\prod_{t=1}^\ell m\left( h, P^t_{\gamma, \pi} \right).
\end{equation}

\medskip
We say that a partition $\gamma\in\Pi (k, j)$ {\em refines} partition
$\sigma\in\Pi(k, \ell)$, $\ell\le j$, if each block of $\gamma$ is contained in
one of the blocks of $\sigma$. In this case, we write $\gamma\prec\sigma$.
Given $1\le\ell\le j\le k$, there is one-to-one correspondence between all possible
pairs of partitions $(\gamma, \pi)$ with $\gamma\in\Pi(k, j)$ and $\pi\in\Pi(j, \ell)$
splitting the set
$\{ \gamma_1, ..., \gamma_j \}$ of blocks of $\gamma$ into $\ell$ `super-blocks', and all possible
pairs of partitions $(\gamma, \sigma)$ with $\gamma\in\Pi(k, j)$ and $\sigma\in\Pi(k, \ell)$ such that
$\gamma \prec \sigma$. We will use the enumeration $\sigma_1, ..., \sigma_\ell$ of blocks of $\sigma$
induced by the enumeration of blocks of $\pi$. For instance,
in the example considered above, $\sigma\in\Pi(8, 2)$, $\sigma_1 = \{1, 2, 3, 4, 5, 6\}$
and $\sigma_2 = \{7, 8\}$.

We denote by  $P^t_{\gamma, \sigma} = P^t_{\gamma, \pi}$, $1\le t \le \ell$,
the set of lengths of blocks of $\gamma$ that are contained in the block $\sigma_t$ of $\sigma$.
Then changing the order of summations in~\eqref{eq-long},
we get
\begin{multline*}
\sum_{\ell=1}^k (-1)^{\ell-1} (\ell-1)! \, \sum_{\sigma\in\Pi(k, \ell)}\,
\sum_{j=\ell}^k \ \sum_{ \substack{\gamma\in\Pi(k, j) \\ \gamma \prec \sigma} } \
\prod_{t=1}^\ell m\left( h, P^t_{\gamma, \sigma} \right) \\
=
\sum_{\ell=1}^k (-1)^{\ell-1} (\ell-1)! \, \sum_{\sigma\in\Pi(k, \ell)}\,
\sum_{ \substack{\gamma\in\Pi(k) \\ \gamma \prec \sigma} } \
\prod_{t=1}^\ell m\left( h, P^t_{\gamma, \sigma} \right).
\end{multline*}

Next, given $\sigma\in\Pi(k, \ell)$, we fix an arbitrary enumeration of
its blocks, put $q_t = |\sigma_t|$, $1\le t\le \ell$,
and replace one partition
$\gamma \prec \sigma $ by $\ell$ partitions $\gamma^1\in\Pi(q_1)$, ..., $\gamma^\ell\in\Pi(q_\ell)$
that split the blocks of $\sigma$ into the corresponding sub-blocks.
In the example considered above, $q_1 = 6$, $q_2= 2$,
the partition $\gamma^1$ consists of two blocks
$\{1, 2, 3 \}$ and $\{ 4, 5, 6\}$, and the partition $\gamma^2$ consists of one block.

At last, replacing the sum
$\displaystyle \sum_{ \substack{\gamma\in\Pi(k) \\ \gamma \prec \sigma} }$
by the $\ell$-tuple sum $\displaystyle \sum_{\gamma^1\in\Pi(q_1)} ... \sum_{\gamma^\ell\in\Pi(q_\ell)} $,
we get
\[
\sum_{ \substack{\gamma\in\Pi(k) \\ \gamma \prec \sigma} } \
\prod_{t=1}^\ell m\left( h, P^t_{\gamma, \sigma} \right)
= \prod_{i=1}^\ell \, \sum_{\lambda\in\Pi(q_i)} \,
m(h, P_\lambda)
\stackrel{\eqref{eq-m_k}}= m_{q_1}(h)\,...\, m_{q_\ell}(h)\,.
\]
Hence, the right-hand side of \eqref{eq-s_k} equals
\[
\sum_{\ell=1}^k (-1)^{\ell-1} (\ell-1)! \, \sum_{\sigma\in\Pi(k, \ell)}
m_{q_1}(h)\,...\, m_{q_\ell}(h) \stackrel{\eqref{eq_mc}}= s_k(h)\,,
\]
proving the claim. \hfill $\Box$

\end{document}